\documentclass[twocolumn,preprintnumbers,amsmath,amssymb]{revtex4}

\usepackage[dvips]{graphicx}

\usepackage{amsmath}
\usepackage{amsbsy}

\usepackage{color}

\DeclareGraphicsExtensions{.pdf,.eps,.png,.jpg,.mps}
\usepackage{gensymb}

\usepackage{xcolor}

\begin{document}

\title{Ultrashort Mn-Mn Bonds in Organometallic Complexes}
\author{ T. Alonso-Lanza (*),$^{1}$ J. W. Gonz\'alez,$^{1}$ F. Aguilera-Granja,$^{1,2}$ A. Ayuela$^{1}$}
\affiliation{
$^1$Centro de F\'{\i}sica de Materiales CFM-MPC CSIC-UPV/EHU, Donostia 
International Physics Center (DIPC), Departamento de F\'{\i}sica de Materiales, Fac. de Qu\'{\i}micas, UPV-EHU, 20018 San Sebasti\'an, Spain
\\
$^2$Instituto de F\'{\i}sica, Universidad Aut\'onoma de San Luis de \\Potos\'{\i}, 78000 San Luis Potos\'{\i} S.L.P., M\'exico
\\
}

\date{\today}

\begin{abstract}
Manganese metallocenes larger than the experimentally produced sandwiched MnBz$_2$ compound are studied using several 
density functional theory methods. First, we show that the lowest energy structures have Mn clusters surrounded by benzene molecules, 
in so-called rice-ball structures. We then find a strikingly short bond length of 1.8 \AA{} between pairs of Mn 
atoms, accompanied by magnetism depletion.  
The ultrashort bond lengths are related to Bz molecules caging a pair of Mn atoms, leading to a Mn-Mn triple bond. This effect is also found when replacing benzenes by other molecules such as borazine or cyclopentadiene.
The stability of the Mn-Mn bond for Mn$_2$Bz$_2$ is further investigated using dissociation energy curves.  
For each spin configuration, the energy versus distance plot shows different spin minima with barriers, 
which must be overcome to synthesize larger Mn-Bz complexes.
\end{abstract}

\maketitle

\section*{Introduction}
A number of organometallic compounds have been synthesized and used in a wide range of applications\cite{crabtree2009organometallic,omae1998applications,kollonitsch1965industrial,coogan2012introduction,harrod2012applications}. 
The  synthesis of ferrocene\cite{kealy1951new,miller1952114,dunitz1956crystal,werner2012least} stimulated a search for related metallocenes, 
such as cobaltocene\cite{liu2007cobaltocene}, nickelocene\cite{pfab1953kristallstruktur}, rhodocene\cite{cotton1953cyclopentadienyl} and 
manganocene\cite{wilkinson1954cyclopentadienyl}, in which a transition metal atom is located between two cyclopentadiene molecules.
Similar compounds have been proposed combining transition-metal atoms with benzene molecules (TM-Bz compounds)\cite{wedderburn2006geometries}.  
The basic unit of such compound is a transition metal atom over a single Bz ring, known as a 
half-sandwich\cite{pandey2001electronic,pandey2000unique,muhida2004spin}. 
Such basic units can be combined to form larger compounds with either sandwich-like structures for early transition metals (Sc, Ti, and V) or rice-ball structures for late transition metals (Fe, Co, and Ni)\cite{kurikawa1999electronic}.
TM-Bz compounds look promising for future applications.
For instance, sandwich-like molecules have been proposed as conductors involving spin transport\cite{ormaza2017efficient}, either isolated bridging 
the tips of a Cu nanocontacts\cite{0953-8984-28-44-445301,ormaza2017efficient} or pile up building infinite wires having Sc, Ti, V, Cr, 
and Mn\cite{xiang2006one}.
Sandwiched molecules and infinite wires containing V, Nb, and Ta have been found to provide strong magnetic 
anisotropy \cite{mokrousov2006magnetic}. 
We are interested in characterizing larger clusters with rice-ball structures, specifically in the study of 
Mn-Bz compounds, to gain knowledge on their still unfamiliar structures and properties.

The field of organometallics was recently boosted following a breakthrough in the synthesis of binuclear metallocenes, so-called dimetallocenes,  
such as decamethyldizincocene\cite{resa2004decamethyldizincocene,grirrane2007zinc} Zn$_{2}$($\eta^{5}$C$_{5}$Me$_{5}$)$_{2}$)
and (Zn$_{2}$($\eta^{5}$C$_{5}$Me$_{4}$Et)$_{2}$). Compounds containing Mn with many nuclei have plenty of uses
in organometallics\cite{cahiez2009chemistry}. Some previously described molecules containing Mn-C bonds are 
dimanganese decacarbonyl\cite{brimm1954preparation,king1968convenient,dahl1957polynuclear} Mn$_{2}$(CO)$_{10}$ and methylcyclopentadienyl 
manganese tricarbonyl (CH$_3$C$_5$H$_4$)Mn(CO)$_{3}$, which has been used to increase the octane level of gasoline\cite{davis1998methylcyclopentadienyl}. 
Experiments on TM-Bz compounds have been performed, leading to their synthesis in the gas phase and their characterization
using spectroscopic methods \cite{duncan2008structures,kurikawa1999electronic,zheng2008photoelectron,buchanan1999novel} and in magnetic
experiments\cite{miyajima2007stern,akin2016size}.  TM-Bz compounds are present in many types of 
molecular clusters. Small Mn-Bz compounds, such as MnBz$_2$, have been synthesized\cite{kurikawa1999electronic,zamudio2015direct}. 
As for Mn adatoms in graphene, Mn sandwiched between Bz molecules prefers to be in a hollow position\cite{sevinccli2008electronic,mao2008density}. Theoretical studies are required to understand larger complexes of inner Mn clusters surrounded by Bz molecules as to be produced experimentally.

\begin{figure}[h!]
\includegraphics[clip,width=0.3\textwidth]{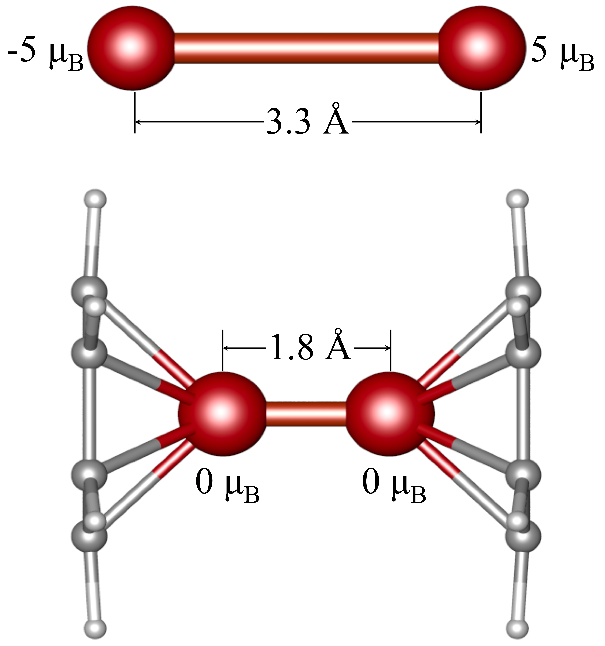}
\caption{\label{figure1} 
Scheme showing the ultrashort bond effect in Mn surrounded by benzene molecules. Note the large reduction in Mn-Mn distance and the total 
quenching of the magnetic moment.} 
\end{figure}

Dimetallocenes and multimetallocenes have been largely investigated using density functional theory (DFT) calculations. Different external molecules have been considered, such as Bz, cyclopentadiene, and even fullerenes\cite{gao2009theoretical}. Similarly, many different central metal atoms have been studied, leading to alkaline-earth 
dimetallocenes\cite{chattaraj2008bonding,kan2009covalent,li2006dft,velazquez2007multimetallocenes}, 
transition metal dimetallocenes\cite{zhou2006novel,zhu2011novel,meng2015two}, heterodinuclear 
compounds\cite{he2008one,hu2015series} and zinc isoelectronic elements, such as Cd and 
Hg \cite{carmona2008direct}. 
Although multinuclear metallocenes of Co, Cu, and Ni, and other transition metals have 
been described
\cite{zhou2006novel,zhu2011novel,meng2015two,rao2002caging,froudakis2001structural,
zhang2008structural,valencia2011rice,kua2006computational,gonzalez2016non}, the study
of larger manganese benzene compounds is still to be done.

We herein investigate organometallic compounds in which Bz molecules cage clusters with a few Mn atoms, Mn$_n$Bz$_m$. 
The benzene molecules strongly affect the bonding between manganese atoms, differently to other transition metals. 
Manganese benzene compounds are also interesting because they are in the middle between two trends, sandwich-like 
structures for early transition metals and rice-ball structures for late transition metals \cite{kurikawa1999electronic}.
We find that the clusters contain Mn-Mn multiple (triple) bonds with ultrashort interatomic distances and depleted 
magnetism. We then focus on the electronic structure of the smaller Mn$_{2}$Bz$_{2}$ 
molecule to further improve our understanding of such molecules. 
The two Bz molecules covering the Mn$_2$ molecule appear to stabilize a non-magnetic local minimum in the inner Mn dimer. 
We verify that these results also appear when replacing the caging benzenes with other molecules.
Furthermore, ultrashort bonds are also found in larger Mn-Bz compounds, indicating that the effect is robust and that strong 
dimerization is induced in larger molecules.
Because most of experiments work with charged clusters, we have checked that positively charged Mn-Bz clusters also present 
ultrashort Mn-Mn bonds. 
Lastly we study the synthesis of the Mn$_{2}$Bz$_{2}$ molecule, as a first step towards larger Mn-Bz compounds, and we show 
that different high local spin magnetic moment states create barriers to such syntheses.

\section{\label{sec:results}Results and discussion}

\subsection{\label{superdim} Metallization induced superdimerization}

\begin{figure*}[thpb]
\centering
\includegraphics[clip,width=0.65\textwidth]{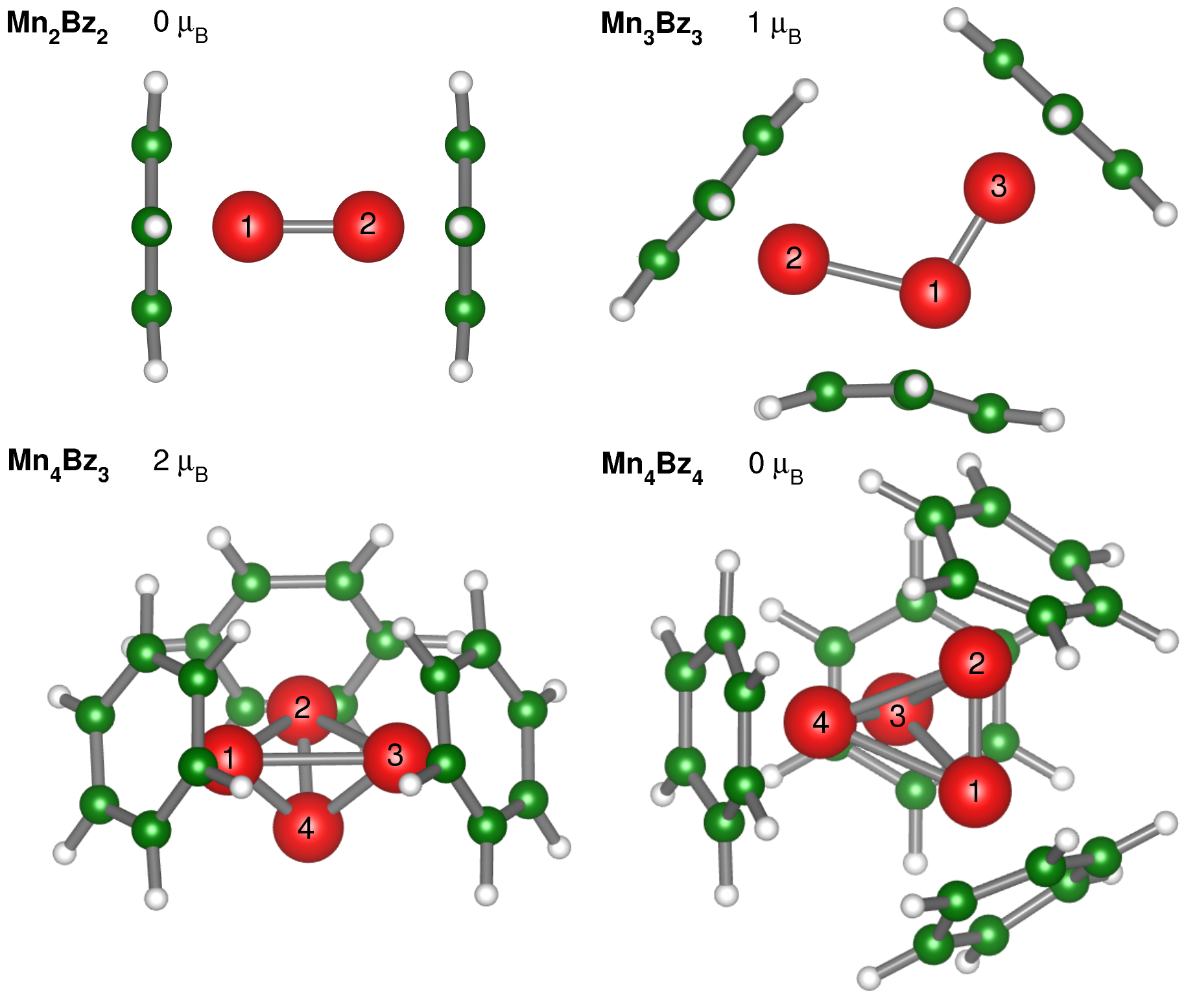}
\caption{\label{figure2} 
Geometries of the Mn$_{m}$Bz$_{n}$ molecules studied here. Mn, C, and H atoms are shown as red, green and white spheres, respectively. Note that the total spin magnetic moment is null or small for all cases.} 
\end{figure*}

The ground state geometries of the Mn$_{2}$Bz$_{2}$, Mn$_{3}$Bz$_{3}$, Mn$_{4}$Bz$_{3}$, and Mn$_{4}$Bz$_{4}$ molecules are shown in Fig. \ref{figure2}. 
We also consider the roles of isomers for each cluster, and study other geometries and magnetic solutions. We first determine whether a sandwich or rice-ball structure is preferred. The rice-ball geometry has lower energy for each of the four Mn-Bz clusters studied here. The results for the sandwich structures are shown in the Supplemental Material. We perform calculations for several isomers with different magnetic moments. For instance, a symmetric Mn$_{3}$Bz$_{3}$ isomer with 5 $\mu_{B}$ has an energy difference of 1 eV from the ground state, and a distorted Mn$_{3}$Bz$_{3}$ isomer with 7 $\mu_{B}$ lies 0.5 eV higher in energy. All the isomers are lying higher in energies than the structures presented in Fig. \ref{figure2}.

All the ground state molecules have strikingly short bond lengths between the Mn couples. 
The Mn$_{2}$Bz$_{2}$ molecule is the simplest system to show this effect with a Mn-Mn distance of 1.8 \AA{}. 
The ultrashort bond lengths are unexpected because larger distances between Mn atoms have been reported, such as 3.4 \AA{} for the Mn dimer\cite{yamamoto2006study,nayak1998equilibrium,baumann1983esr}, 
2.93 \AA{} for dimanganese decacarbonyl\cite{brimm1954preparation,dahl1957polynuclear}, 2.4--2.6 \AA{} for manganese carbides\cite{chong2014first,karen1991phase} and the broad distance range of 
2.25--2.95 \AA{} for several Mn bulk allotropes \cite{baumann1983esr}.
On the other hand, analogous short multiple bonds have been shown theoretically and experimentally for other 
transition metals, such as in ultrashort Cr dimers (1.73 \AA{}) within organic molecules \cite{wagner2009ultrashort}.

We focus on the locally unpolarized Mn$_{2}$Bz$_{2}$ molecule with a Mn-Mn distance of around 1.8 \AA{}. 
The loss of magnetism is a hallmark of the ultrashort bonds between Mn atoms capped with Bz molecules. 
The magnetic moments of irregular systems, such as Mn$_{3}$Bz$_{3}$ and Mn$_{4}$Bz$_{3}$, are almost depleted. 
In the unit consisting of a Mn atom over a benzene molecule, the magnetic moment decreases from 5 $\mu_{B}$ to 
3 $\mu_{B}$ \cite{pandey2000unique}. However, the total elimination of the magnetism for the Mn$_{2}$Bz$_{2}$ molecule 
is still surprising. 

When combining two Mn$_{2}$Bz$_{2}$ units to form a Mn$_{4}$Bz$_{4}$ molecule, the larger compound also has ultrashort bonds, and the total and 
local magnetic moments remain zero. The Bz molecules around the Mn atoms form a distorted tetrahedron, tilted because of neighboring 
Bz rings. There are two short bond lengths of 1.94 \AA{} and four larger bond lengths of 2.56 \AA{}. 
The short bond lengths are between the opposite atom pairs, labeled 1-2 and 3-4 in Fig. \ref{figure2}.

The two cases Mn$_{2}$Bz$_{2}$ and Mn$_{4}$Bz$_{4}$ have ultrashort distances and total magnetism depletion. 
The ultrashort Mn-Mn bonds also occurs for other molecules, such as  Mn$_{3}$Bz$_{3}$ and Mn$_{4}$Bz$_{3}$.
The Mn$_{3}$Bz$_{3}$ ground state geometry depletes most of the magnetism, and gives an ultrashort distance of 1.88 \AA{} between 
Mn(1) and Mn(3). The third Mn atom, Mn(2), moves away and remains bonded mainly to Mn(1), with a bond length of 2.22 \AA{}. 
The side position of Mn(2) bends the Bz molecule associated with Mn(1). 
The total spin magnetic moment is 1 $\mu_{B}$, localized mainly on Mn(2) with 1.53 $\mu_{B}$. 
The Mn(1) atom has antiferromagnetic coupling with a magnetic moment of -0.57 $\mu_{B}$, 
and the Mn(3) local moment decreases to a very small value of 0.28 $\mu_{B}$. \footnote{Small spin polarization is 
induced in the C atoms (0.06 $\mu_{B}$ being the highest value for one C atom) with opposite sign to the closest Mn atom.} 
The added Mn-Bz unit distorts the Mn$_{2}$Bz$_{2}$ geometry, giving nonzero local magnetic moments on the Mn atoms.
The peculiar distorted structure is due to the large energy gain because of the ultrashort Mn-Mn bond. 
The ultrashort bond phenomenon is therefore found partially for Mn$_{3}$Bz$_{3}$.
The dimerized Mn$_{4}$Bz$_{4}$ and Mn$_{3}$Bz$_{3}$ structures indicate that the ultrashort bond mechanism is caused by 
two Mn atoms caged by Bz molecules.

We next consider when the short Mn-Mn bonds accompanied by magnetism quenching occur by 
removing a Bz molecule from an already dimerized structure. We consider Mn$_{4}$Bz$_{3}$ molecule shown in 
Fig. \ref{figure2}. The structure is obtained by removing a Bz molecule from the final Mn$_{4}$Bz$_{4}$ geometry, 
and fully relaxing the geometry again. 
The  Mn$_{4}$Bz$_{3}$ molecule becomes reorganized because the Mn(4) atom has lost its Bz molecule. The Mn(1), Mn(2) and Mn(3) atoms form an equilateral 
triangle with sides 2.62 \AA{}. The Mn(4) atom is bonded to the other three Mn atoms with short bond lengths of 2.05 \AA{}, 
which are larger than those found in Mn$_{4}$Bz$_{4}$. 
In this case, the local magnetic moments are antiferromagnetically coupled, with small values of 1.07 $\mu_{B}$ for Mn(1), Mn(2) and Mn(3) 
and -0.83 $\mu_{B}$ for Mn(4). Most of carbon atoms have a negative induced magnetic moment of 0.02--0.03 $\mu_{B}$. 
Although the local manganese magnetism is reduced and the bonds are short, it seems that the presence of two Bz molecules 
caging two Mn atoms is key to the ultrashort Mn-Mn bond scheme.

We now consider the stability these compounds to discuss their possible synthesis. 
We compute the formation energy, per manganese atom, for Mn$_{n}$Bz$_{m}$ clusters toward separated 
Mn atoms and Bz molecules, using the expression $E_f = (n \, E_{\mathrm{Mn}} + m \, E_{\mathrm{Bz}} - E_{\mathrm{total}})/n$. 
We find that the formation energies are always favorable, in the range 1.76--2.18 eV.
Furthermore, the HOMO-LUMO gaps are in the range 1.17--0.72 eV. Taking into account the well-known issue of gap 
underestimation for generalized gradient approximation, the gap values suggest that the molecules are stable.  
The gap narrows as the size of the molecule increases. 
We could have followed the series by looking for larger molecules with ultrashort Mn-Mn bonds, 
such as Mn$_{5}$Bz$_{5}$, Mn$_{6}$Bz$_{6}$, and so on. Nevertheless, it appears that placing five or more Bz 
rings around a nucleus of larger Mn clusters is a real challenge due to the lack of space. 

Molecular beam experiments use charged systems, therefore we determine how our neutral 
Mn-Bz clusters are affected when positively charged. An electron was removed from the structures, which were fully 
relaxed again. The results show there are two different trends depending on the size. 
Smaller clusters, Mn$_{2}$Bz$_{2}^{+}$ and Mn$_{3}$Bz$_{3}^{+}$ preserve the ultrashort distance of the Mn-Mn bonds. 
Note that the Mn$_{2}$Bz$_{2}^{+}$ cluster now has a magnetic moment of 1 $\mu_{B}$ and 
that the Mn$_{3}$Bz$_{3}^{+}$ increases from 1 $\mu_{B}$ for the neutral case to 2 $\mu_{B}$.
Larger complexes, Mn$_{4}$Bz$_{3}^{+}$ and Mn$_{4}$Bz$_{4}^{+}$ lost the ultrashort Mn-Mn bond distance and display bulk-like behavior. 
The Mn-Mn distances increases to values within the interval 2.3--2.6 \AA{}. The local magnetic moments of Mn increase to values 
within the range 2.0 -- 3.5 $\mu_{B}$ and the Mn atoms are antiferromagnetically coupled.

We further test when ultrashort Mn-Mn bonds would appear by replacing the caging benzenes with other organic molecules, such as cyclopentadiene C$_5$H$_6$,
 and even inorganic molecules, such as borazine B$_3$H$_6$N$_3$ \cite{zhu2009ab,li2013magnetic,wilkinson1956manganese} .
After careful atomic relaxations, both types of caging compounds for Mn$_2$ present ultrashort Mn-Mn bonds accompanied by magnetism depletion. \footnote{For the cyclopentadiene-caged Mn$_2$, the ultrashort Mn-Mn distance is 1.85 \AA{}. The Mn local magnetic moments become about 0.1 $\mu_B$, which are not fully depleted in comparison to Bz-Mn molecule because a H atom from cyclopentadiene is bound to a Mn atom. For the borazine-caged Mn$_2$, the Mn-Mn distance is 1.82 \AA{} and the local and total magnetic moments are fully depleted.}
Then, the ultrashort Mn-Mn bond effect, accompanied by a significant decrease in the magnetic moment of Mn, seems to be robust, and it is expected to be found for other molecules caging manganese atoms.

\subsection{\label{sec:dimer} Simplest case: Mn$_{2}$Bz$_{2}$}
We focus on the smaller Mn$_{2}$Bz$_{2}$ molecule to investigate the ultrashort Mn-Mn bond, also found in larger molecules.
The fully relaxed Mn$_{2}$Bz$_{2}$ molecule has an unexpected Mn-Mn bond length of 
1.8 \AA{}, as shown in Fig. \ref{figure2}.
However, the Mn dimer has been found to be stabilized by van der Waals interactions \cite{yamamoto2006study} and to have a bond length of more than 
3 \AA{}. Bulk metallic Mn is predicted to have interatomic distances of between 2.25 \AA{} and 2.95 \AA{}, depending on the allotropic structure.
The Mn-Mn distance decreases when shifting from van der Waals-like bonds to metallic-like bonds. However, our results are surprising because 
the Mn-Mn bonds can be ultrashort. This means that the Mn-Mn bonding mechanism in hybrid Mn-Bz molecules is different from the one seen in 
Mn clusters and bulk Mn \cite{nayak1998equilibrium}.  
We find a non-magnetic solution for the Mn$_{2}$Bz$_{2}$ molecule that contrasts with the high-spin solutions 
obtained from most DFT calculations for the Mn
dimer \cite{qing2007configurations,bobadova2005structure,longo2005fully,kabir2006structure,nayak1998equilibrium,baumann1983esr}.
Benzene molecules reduce spin-splitting in the Mn atoms, and electrons fill the minority-spin bonding molecular orbitals instead of 
the majority-spin antibonding molecular orbitals. 
The ultrashort Mn-Mn distance is therefore related to the loss of magnetism. For spin-compensated solutions, the Mn dimer 
in Mn$_{2}$Bz$_{2}$ can be further compressed because electrons between the Mn atoms suffer less electronic repulsion when 
they fill the same Mn-Mn molecular levels with opposite spins. Note that high-spin solutions for Mn$_{2}$Bz$_{2}$ require 
larger distances because of the larger electronic repulsion between Mn levels.

\begin{figure}[thpb]
\includegraphics[clip,width=0.4\textwidth]{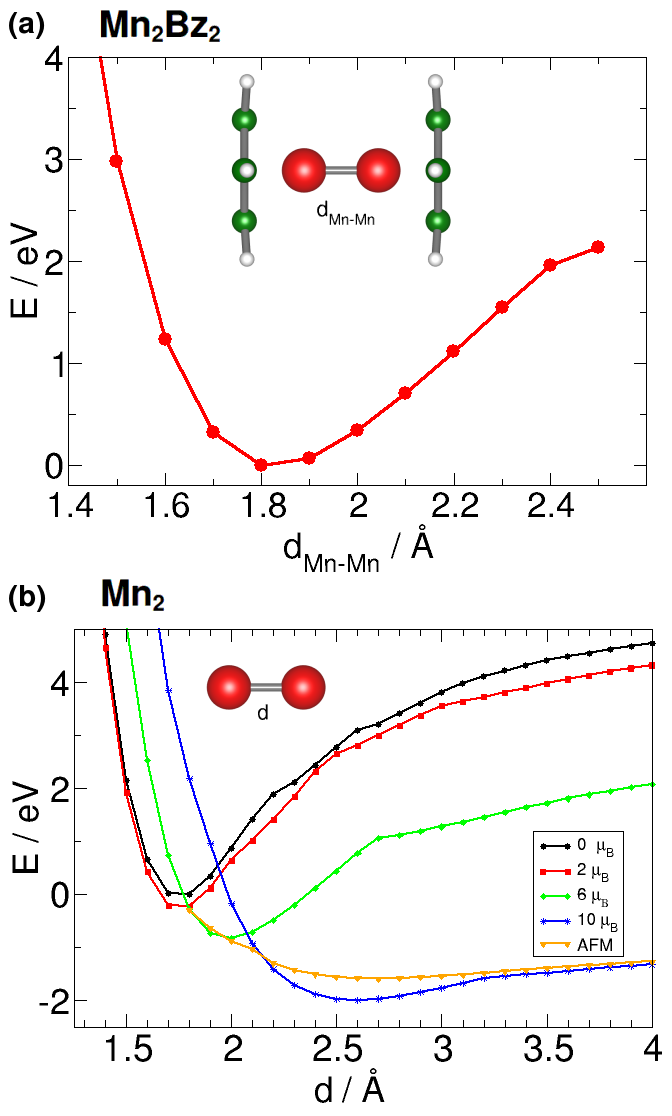}
\caption{\label{figure3} 
(a) Total energy versus Mn-Mn bond length for the Mn$_{2}$Bz$_{2}$ cluster.
(b) Total energy versus distance for the Mn$_{2}$ dimer. Each curve  has  the total magnetic moment fixed at a different value.}
\end{figure}

To improve our understanding of the ultrashort Mn-Mn bonds, we carry out spin-polarized single point calculations for the Mn$_{2}$Bz$_{2}$ 
molecule. The Mn atoms are separated by 0.1 \AA{} in each step, and the Bz rings are maintained in the same relative positions with 
respect to the Mn atoms. We plot the energy against Mn-Mn distance in Fig. \ref{figure3}(a).
The local magnetic moments of Mn are zero up to a distance of 2.4 \AA{}, and there is a deep minimum at a distance of 1.8 \AA{}.  
The energy increases sharply by more than 2 eV as the Mn atoms separate. Two different DFT codes gives the minimum at 1.8 \AA{},
even when the Mn$_{2}$Bz$_{2}$ molecule is distorted and allowed to relax again.
The large energy gain is associated to the multiple (triple) bond between Mn atoms, as shown by considering the molecular orbitals and electron localization function, as described in the Supplemental Material.

We further explore the nature of Mn$_{2}$Bz$_{2}$ bonding to determine when both inner Mn atoms are locally spin-compensated.
We study the isolated Mn$_{2}$ molecule, varying the Mn-Mn distance from 1 \AA{} to 4 \AA{}, as shown in Fig. \ref{figure3}(b).
When the magnetization is released, the total magnetic moment varies from 2 $\mu_B$ for distances shorter than 1.7 \AA{} to 10 $\mu_B$ 
for distances larger than 2.4 \AA{}.
We have also calculated the energy versus distance curves for fixed total spin values of 0, 2, 6, and 10 $\mu_B$.  
We find that each spin curve has a different minimum.
The global minimum for Mn$_{2}$ is around 2.5 \AA{}, as previously reported at this level of 
calculations (LDA/GGA)\cite{qing2007configurations,bobadova2005structure,longo2005fully,kabir2006structure}. 
The antiferromagnetic curve is almost degenerated to that of 10 $\mu_B$ for long distances, in agreement with previous 
calculations\cite{tzeli2008first}$^,$\footnote{There is a delicate balance between exchange and splitting, and small changes 
in distance can modify the magnetic characteristics of the Mn dimer\cite{mejia2008understanding}. Although DFT has been shown 
to be successful for many systems, the choice of exchange correlation functionals is key for some systems, such as the Mn dimer. 
Note, for example, that obtaining the minimum caused by van der Waals interactions\cite{morse1986clusters,nayak1998equilibrium,lopez2003magnetism} 
at distances larger than 3 \AA{} requires using hybrid functionals\cite{yamanaka2007density} or even the \textsc{casscf} 
(Complete Active Space Self-Consistent Field) method\cite{yamamoto2006study}, which reproduces the antiferromagnetic coupling and 
the bond length larger than 3 \AA{} obtained in experiments\cite{cheeseman1990transition,baumann1983esr,rivoal1982ground,
van1981antiferromagnetic,bier1988resonance}.}.
The strong Mn-Mn bonding in the Mn$_{2}$ molecule at short distances can therefore be computed accurately enough using 
generalized gradient approximation (GGA) functionals.
The low spin solutions (with 0 and 2 $\mu_B$) shrink to local minima at bond lengths of about 1.8 \AA{}, as reported 
previously \cite{wolf1980nonempirical,paul2009magnetic}. The minimum at short distances is also favored in the bonding in Mn$_{2}$Bz$_{2}$ 
because the magnetic moment in the 3\textit{d} Mn levels is decreased by the Bz molecules, so the 
Mn-Mn distances are dramatically reduced.

\subsection{\label{sec:synthesis} Implications for the synthesis of Mn$_{2}$Bz$_{2}$}

The MnBz$_{2}$ molecule has been synthesized and identified using spectroscopic techniques in the same way as for other TM-Bz compounds \cite{kurikawa1999electronic}. Larger Mn-Bz compounds are, however, difficult to synthesized\cite{yasuike1997vanadium}. It seems that 
electron spin restrictions may prevent their synthesis.
Here, we investigate the path for the synthesis of Mn$_{2}$Bz$_{2}$  by estimating energy barriers. 
The possible barriers are extracted from the energy versus Bz-Bz distance curves for different total magnetic moments, 
as shown in Fig. \ref{figure4}. 
Further computational details are given in the Supplemental Material. 
The total magnetic moment decreases from long to short distances, and a barrier appears as the 
Mn$_{2}$Bz$_{2}$ molecule forms at high spin. We  can overcome this barrier by, for instance, bringing the Mn-Mn 
core into a low spin state sandwiched between distant Bz molecules. 
We propose that using thermal or optical pumping techniques on the Mn cluster and Bz 
molecule precursors could help to find routes for synthesizing the Mn$_{2}$Bz$_{2}$ molecule and larger Mn-Bz clusters.

\begin{figure}[thpb]
\includegraphics[clip,width=0.4\textwidth]{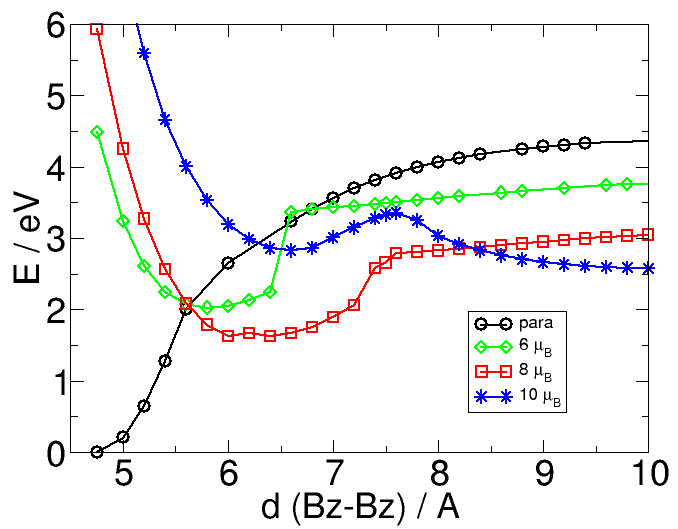}
\caption{\label{figure4} 
Mn$_{2}$Bz$_{2}$ total energies for different distances between the benzene molecules. The curves were obtained by fixing the total magnetic moment.} 
\end{figure}

\section{\label{sec:sum} Final remarks}
We studied Mn-Bz molecules with rice-ball configurations. We found a multiple (triple) bond between the inner manganese atoms in the core metal cluster. 
Two hallmarks of this effect are a strikingly short Mn-Mn bond length of about 1.8 \AA{} and a lack of magnetism. These can be explained because every two Bz molecules stabilize the two neighboring Mn atoms with no spin polarization. Furthermore, other molecules such as borazine and cyclopentadiene have a similar effect on manganese atoms.
We also found that the ground states of larger molecules, such as Mn$_{3}$Bz$_{3}$ and Mn$_{4}$Bz$_{4}$, stabilize the irregular structures 
of the molecules with ultrashort Mn-Mn bonds. 
Last, we investigated the roles of different spins, looking at the Mn$_{2}$Bz$_{2}$ energy as the Bz-Bz distance varied. 
There are energy barriers that could be overcome by canceling local Mn spins using optical or thermal techniques.  
These results suggest that further experimental progress can be made in synthesizing larger Mn-Bz clusters that are of interest in the field of organometallic compounds.

\section{\label{sec:compdetails} Computational details}

We performed DFT calculations using the Vienna ab-initio simulation package (\textsc{vasp}), based on the projected augmented wave method
(PAW)\cite{blochl1994projector,kresse1999ultrasoft}. For the exchange and correlation functionals we used the Perdew-Burke-Ernzenhof (PBE)
form of the generalized gradient approximation (GGA)\cite{perdew1996generalized}.
We chose a cubic unit cell with sides 30 \AA{} in order to avoid interaction between images.
We used an electronic temperature of 25 meV and a cutoff energy of 500 eV.
The Mn-Bz molecules were relaxed until the forces were less than 0.006 eV/\AA{}.
The results obtained using \textsc{vasp} were reproduced using the \textsc{siesta} package within a GGA and applying the PBE functional for selected cases.
We used a mesh cut-off of 250 Ry and sampled at the $\Gamma$-point.
The atomic cores were described using nonlocal norm-conserving relativistic Troullier-Martins\cite{troullier1991efficient} 
pseudopotentials with non-linear core corrections factorized in the Kleynman-Bylander form. 
The basis set size was double zeta plus polarization orbitals with an energy shift of 0.05 eV.
Using the \textsc{siesta} package, the same minima were obtained for the Mn$_{2}$Bz$_{2}$
and Mn$_{3}$Bz$_{3}$ molecules.

\section*{Acknowledgments}
This work was partially supported through projects FIS2013-48286-C02-01-P and FIS2016-76617-P funded by the Spanish Ministry of Economy and Competitiveness MINECO, by the Basque Government under the ELKARTEK project(SUPER), and by the University of the Basque Country  (Grant No. IT-756-13). TA-L acknowledges a grant provided by the MPC Material Physics Center - San Sebasti\'an. 
FA-G acknowledges the DIPC for their generous hospitality. 
The authors also acknowledge the support of the DIPC computer center.

\clearpage

\begin{center}
\textbf{\large Supplemental Material: Ultrashort Mn-Mn Bonds in Organometallic Complexes}
\end{center}
\setcounter{figure}{0} 
\setcounter{section}{0} 
\setcounter{equation}{0}
\setcounter{page}{1}
\renewcommand{\thepage}{S\arabic{page}} 
\renewcommand{\thesection}{S\Roman{section}}   
\renewcommand{\thetable}{S\arabic{table}}  
\renewcommand{\thefigure}{S\arabic{figure}} 
\renewcommand{\theequation}{S\arabic{equation}} 

\section{\label{sec:Appendix 1} M\MakeLowercase{n}-M\MakeLowercase{n} bond in M\MakeLowercase{n}$_{2}$B\MakeLowercase{z}$_{2}$}

\subsection{\label{sec:orbitals} Molecular orbitals}

We studied the hybridization of the six $p_z$ molecular orbitals of benzene (Bz) and the 3\textit{d} shell of Mn.
The properties of benzene orbitals are well known. Three of the four valence electrons of each carbon atom are used to build 
strong $\sigma$ bonds with neighboring C and H  atoms. The extra electrons in the p$_{z}$ orbitals 
of C atoms form delocalized $\pi$ bonds, which are related to the aromaticity of Bz. These p$_{z}$ electrons are 
higher in energy than $\sigma$ electrons and close to the HOMO level, and are responsible for the reactivity of Bz 
with other atoms and molecules. The six p$_{z}$ orbitals are involved in the six well-known molecular orbitals. 
Some molecular orbitals are degenerated and are divided into four types: $\sigma$ ($\pi_{1}$), $\pi$ ($\pi_{2}$,$\pi_{3}$), 
$\delta$ ($\pi_{4}^{*}$,$\pi_{5}^{*}$) and $\phi$ ($\pi_{6}^{*}$).
For Mn$_{2}$Bz$_{2}$, the degenerate Bz levels are, however, split in energy because the Bz molecules are 
slightly bent toward the Mn atoms.

\begin{figure}[thpb]
\includegraphics[clip,width=0.5\textwidth,angle=0,clip]{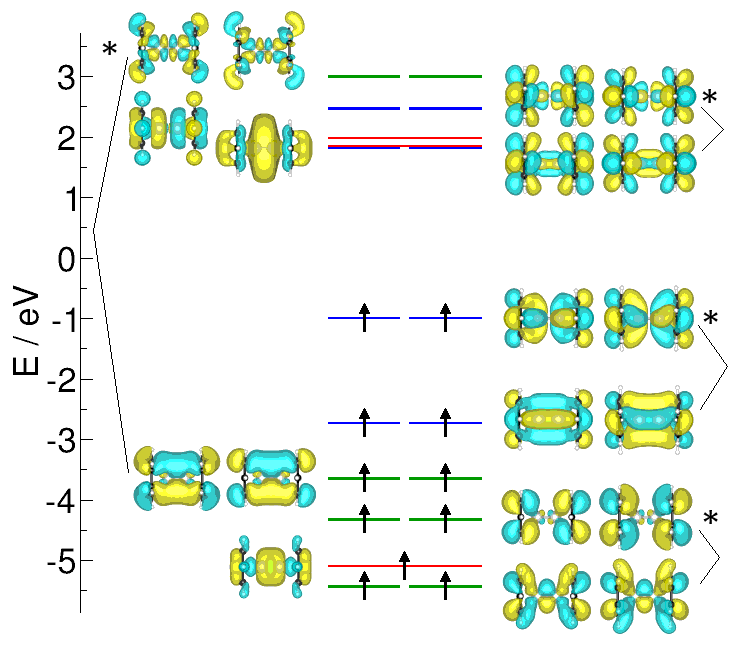}
\caption{\label{figure5} 
Energy levels and wave-functions near the Fermi energy. The colors of the levels denote the molecular orbital types: $\sigma$ in red,  $\pi$ in 
green, and $\delta$ in blue. Antibonding molecular orbitals are marked with  asterisks.} 
\end{figure}

In the presence of a Bz molecule, the Mn 3\textit{d} orbitals  split into three different energies, according to the orbital symmetry, 
in the energy order d$\sigma$ (3\textit{d}$_{z^{2}}$), d$\pi$ (3\textit{d}$_{xz}$,3\textit{d}$_{yz}$), and 
d$\delta$ (3\textit{d}$_{xy}$,3\textit{d}$_{x^{2}-y^{2}}$).
The Mn orbitals hybridize with the Bz orbitals when they have the same symmetry, to produce $\sigma$, $\pi$ and $\delta$ 
molecular orbitals, as has been found in similar analyses \cite{yasuike1999ionization,asd}.
The Mn 3\textit{d}$_{z^{2}}$ orbitals are not expected to interact much with the Bz because they point toward the 
center of the Bz ring. Regarding the Mn 4s orbitals, $\sigma$ bonds with the Bz $\pi_{1}$ molecular orbitals are expected.

The energy levels for Mn$_{2}$Bz$_{2}$ near the Fermi energy are shown in Fig. \ref{figure5}. 
The $\sigma$, $\pi$, and $\delta$ molecular orbitals are found by considering the number of nodal planes, (0, 1, and 2, respectively). 
The two $\pi$ orbitals have lower energies than the $\delta$ orbitals, and are occupied by more electrons. 
Figure \ref{figure5} shows that the $\pi$ and $\delta$ molecular orbitals have bonding or antibonding character that alternates as 
the energy energy increases.
The number of hybrid orbitals expected in each symmetry can also be explained. For $\pi$ symmetry, there are four $\pi$ orbitals 
from two Bz molecules and four 3\textit{d} orbitals (3\textit{d}$_{xz}$ and 3\textit{d}$_{yz}$ from each atom). As a result, 
there are eight molecular orbitals. The same argument can be applied to the $\delta$ orbitals.

To summarize, a spin compensated structure was found with electrons filling the $\sigma$, $\pi$ and $\delta$ hybrid molecular orbitals 
built from the Bz p$_z$ levels and Mn 3\textit{d} levels, following symmetry considerations. Therefore, the ultrashort Mn-Mn bond 
is a triple bond, with $\sigma$, $\pi$ and $\delta$ bonds.

\subsection{\label{sec:elf} Electron localization function }

\begin{figure}[thpb]
\includegraphics[clip,width=0.5\textwidth,angle=0,clip]{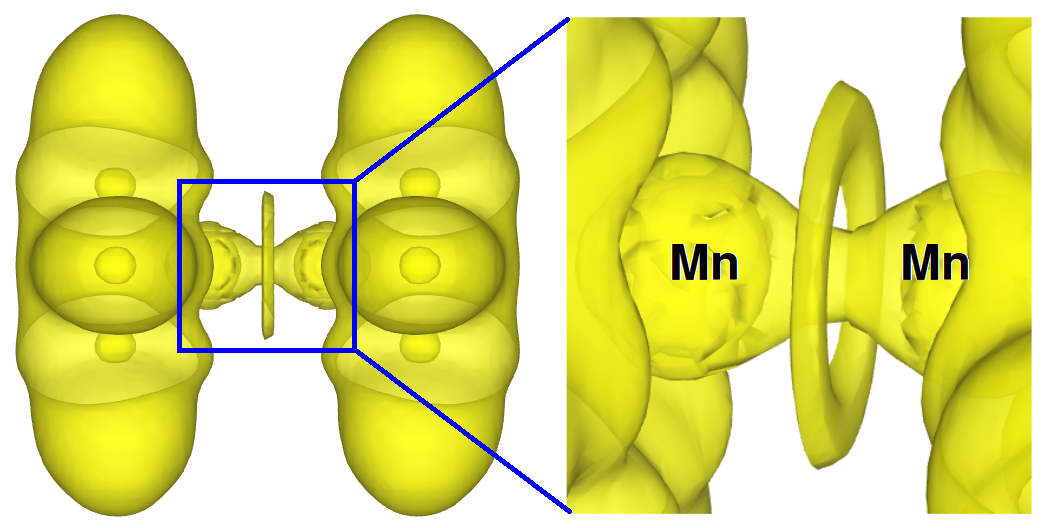}
\caption{\label{figure6} 
(Left) Electron localization function for the Mn$_{2}$Bz$_{2}$ molecule cut at a value of 0.125. 
(Right) An expanded view of the highlighted part of the left figure.} 
\end{figure}

The Mn-Mn bond is studied in detail using the electron localization function, as shown in Fig. \ref{figure6}.
There is a direct $\sigma$ metallic bond between the Mn atoms, composed of 3\textit{d}$_{z^{2}}$ orbitals. The disk around the bond center 
indicates the presence of $\pi$ and $\delta$ molecular orbitals and bonds.
The basins in the Mn-Mn bonds have an electron localization function value of 0.125, which is small and typical of metallic bonds. 
The electron localization function results agree completely with the energy levels, hybridization, and wave-functions described above, 
confirming the triple-bond characteristics of the bonds between the Mn atoms.

\section{\label{sec:Appendix 2} Energy barriers}

\begin{figure}[h]
\includegraphics[clip,width=0.45\textwidth,angle=0,clip]{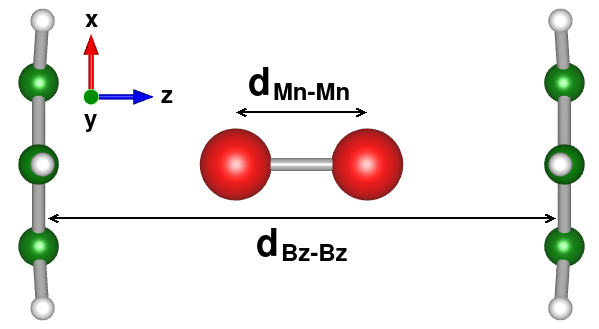}
\caption{\label{figure7} 
 Scheme of the geometry used for the barrier calculations.} 
\end{figure}

\begin{figure}[h]
      \centering
\includegraphics[clip,width=0.45\textwidth,angle=0,clip]{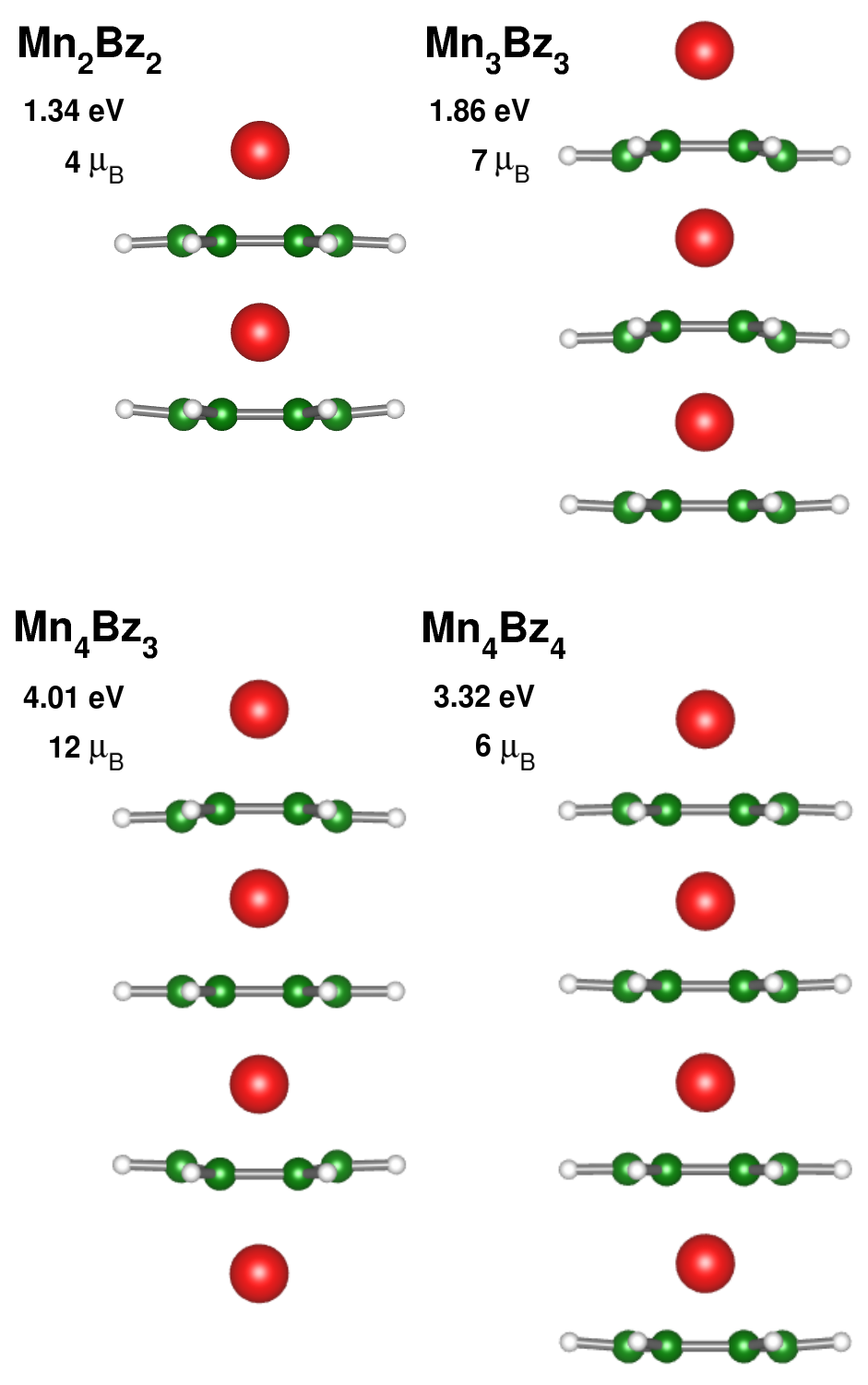}
\caption{\label{figure8} 
Sandwich-like isomers of the four different Mn-benzene clusters studied. The energy difference with respect to the corresponding rice-ball structure and the total magnetic moment for each cluster are included.} 
\end{figure}

We next comment on the barrier that prevents the formation of the Mn$_{2}$Bz$_{2}$ molecule. We carried out calculations for Mn$_{2}$Bz$_{2}$ for a set of spin magnetic moments and distances.
The energy barrier is calculated using the geometrical arrangement shown in Fig. \ref{figure7}. 
The total magnetic moment is fixed to 0, 6, 8, and 10 $\mu_B$. For each spin value, the molecule is computed using several 
Bz-Bz distances d$_{\mathrm{Bz-Bz}}$ between 4.8 and 10 \AA{} at steps of 0.2 \AA{}. 
The C and H atoms are relaxed in the $xy$ plane, keeping the Bz-Bz distance frozen. 
For each calculation, the Mn atoms are initially set in the center with a distance d$_{\mathrm{Mn-Mn}}$ of 2 \AA{}, 
as shown in Fig. \ref{figure7}. 
The Mn atoms are allowed to relax along the $z$ axis. For large d$_{\mathrm{Bz-Bz}}$ values, the Mn atoms can either bind 
to Bz molecules or form a Mn-Mn core in the center. These two geometries explain the abrupt step seen for some 
energy curves in Fig. 4 in the main article, for instance at 6.4 \AA{} for 6 $\mu_B$, and at 7.2 \AA{} for 8 $\mu_B$. 
At these distances, the Mn-Mn forces are at maxima, and for larger distances the Mn-Mn atoms prefer to bond 
together remaining separated from the Bz molecules.

\section{\label{sec:Appendix 3} M\MakeLowercase{n}-B\MakeLowercase{z} sandwiches}

We also considered sandwiched structures for the four Mn-Bz clusters, as shown in Fig. \ref{figure8}. 
The rice-ball structures presented in the main text are more stable than the sandwiched structures, 
with energy differences larger than 1 eV. Furthermore, the total magnetic moments are larger for the sandwiched structures than for the rice-ball structures.

\section{\label{sec:Appendix 4} M\MakeLowercase{n}-B\MakeLowercase{z}: charge state}

We studied local charges using the Mulliken scheme within the \textsc{siesta} package. For the Mn$_{2}$Bz$_{2}$ molecule, there is a small charge transfer of 0.18 electrons from each Mn atom towards C atoms. For Mn$_{3}$Bz$_{3}$, the charge transfers of Mn(1) and Mn(2) are almost negligible, and the charge transfer reaches a threshold of 0.11 electrons for Mn(3). For Mn$_{4}$Bz$_{3}$, Mn(1), Mn(2), and Mn(3) lose 0.07 electrons each, and Mn(4) gains 0.42 electrons; globally each Mn atom therefore receives 0.21 electrons from Bz molecules. For Mn$_{4}$Bz$_{4}$, each Mn atom gains almost 0.04 electrons, meaning that, in total,  0.16 electrons move from Bz molecules to Mn atoms. We therefore find that increasing the Mn cluster size causes the charge transfer between Mn atoms and Bz molecules to change sign.
In general, charge transfer values are small. We can assume that the Mn oxidation state is zero for the calculated Mn-Bz clusters, and specifically for the ultrashort Mn-Mn bond in Mn$_{2}$Bz$_{2}$.


\end{document}